\documentclass[conference]{IEEEtran}
\IEEEoverridecommandlockouts
\usepackage{cite}
\usepackage{amsmath,amssymb,amsfonts}
\usepackage{algorithmic}
\usepackage{graphicx}
\usepackage{textcomp}
\usepackage[table]{xcolor}
\def\BibTeX{{\rm B\kern-.05em{\sc i\kern-.025em b}\kern-.08em
    T\kern-.1667em\lower.7ex\hbox{E}\kern-.125emX}}
    
\usepackage{balance}
\usepackage{xspace}
\usepackage{bm}
\usepackage[labelfont={bf,small},textfont={bf,small}]{caption} 
\usepackage[labelfont=bf,textfont=footnotesize]{subcaption}

\usepackage{changes}

\definecolor{Gray}{gray}{0.9}

\usepackage{pifont} 
\newcommand{\cmark}{\ding{51}}%
\newcommand{\xmark}{\ding{55}}%

\newcommand{\sysname}{MD\-Inference\xspace}%

\newcommand{\thetitle}{\sysname: Balancing Inference Accuracy and Latency for Mobile Applications}

\newcommand{\framework}{framework\xspace}
\newcommand{\frameworks}{frameworks\xspace}
\newcommand{\accuracy}{aggregate accuracy\xspace}
\newcommand{\Accuracy}{Aggregate accuracy\xspace}

\graphicspath{ {figures/} }

\newcommand{\para}[1]{\textbf{#1}}

\begin{document}

\title{\thetitle
}

\author{
  \IEEEauthorblockN{Samuel S. Ogden}
  \IEEEauthorblockA{\textit{Worcester Polytechnic Institute}\\
  ssogden@wpi.edu}
  \and
  \IEEEauthorblockN{Tian Guo}
  \IEEEauthorblockA{\textit{Worcester Polytechnic Institute}\\
  tian@wpi.edu}
}

\maketitle
\thispagestyle{plain}
\pagestyle{plain}

\begin{abstract}
Deep Neural Networks are allowing mobile devices to incorporate a wide range of features into user applications.
However, the computational complexity of these models makes it difficult to run them effectively on resource-constrained mobile devices.
Prior work approached the problem of supporting deep learning in mobile applications by either decreasing model complexity or utilizing powerful cloud servers.
These approaches each only focus on a single aspect of mobile inference and thus they often sacrifice overall performance.

In this work we introduce a holistic approach to designing mobile deep inference \frameworks.
We first identify the key goals of \emph{accuracy} and \emph{latency} for mobile deep inference and the conditions that must be met to achieve them.
We demonstrate our holistic approach through the design of a hypothetical \framework called \sysname.
This \framework leverages two complementary techniques; a model selection algorithm that chooses from a set of cloud-based deep learning models to improve inference accuracy and an on-device request duplication mechanism to bound latency.
Through empirically-driven simulations we show that \sysname improves aggregate accuracy over static approaches by over 40\% without incurring SLA violations.
Additionally, we show that with a target latency of 250ms, \sysname increased the aggregate accuracy in 99.74\% cases on faster university networks and 96.84\% cases on residential networks.

\end{abstract}


\IEEEpeerreviewmaketitle

\section{Introduction}
\label{sec:intro}

Deep learning on mobile devices is allowing for a wide range of new features such as virtual personal assistants~\cite{capes2017siri, kepuska2018next}, visual text translation~\cite{johnson2016googles} and facial filters~\cite{snapchat} to become commonplace in mobile applications.
These diverse functionalities are made possible by recent advanced in machine learning models called \textit{deep neural networks} (DNNs), which on some tasks can approach human-level accuracy~\cite{He2015a}.

However, DNNs achieve this high accuracy with high computational complexity~\cite{Zoph2018} leading to high latency, especially when running on mobile devices~\cite{tensorflow:model-zoo}.
This causes a necessary trade-off to be made between model accuracy and model execution latency.
Modern frameworks such as TensorFlow allow for on-device execution, in-cloud execution, or some hybrid of these two, introducing a wide range of choices for this accuracy-latency trade-off.

Each of these three approaches each have strengths but introduce additional drawbacks.
\emph{On-device inference} allows for executing inferences entirely on the mobile device with easy to predict latency but the mobile developer has to choose between high execution latency or using lower accuracy models.
\emph{In-cloud inference} can execute high-accuracy models with low latency but the reliance on network communication means unpredictable, and potentially unacceptably long, overall response time~\cite{Netravali2015}.
\emph{Hybrid inference} involves spreading execution between the mobile device and the cloud allowing for potential reductions in latency, but can result in worse latency and lower accuracy than purely on-device or in-cloud approaches.

In this paper we argue the need for mobile-oriented inference frameworks.
We discuss the pros and cons of existing approaches and pinpoint the potential areas for improvement.
We propose a holistic approach that considers mobile-specific factors when designing mobile inference \frameworks. 
Finally, we demonstrate our approach through the design of a hypothetical
\framework called \sysname aiming to increase \emph{\accuracy}, defined as the
average accuracy for all models used to service requests, while bounding latency
for mobile inference requests. 
This is enabled by both utilizing a network-aware model selection algorithm to dynamically choose high-accuracy models that can execute within a target response time and duplicating requests to ensure a bounded latency response.

Instead of approaching the design of mobile inference \frameworks as a static problem, where a single model is used and network time is disregarded, we consider a run-time approach to mobile inferences with a two-pronged design. 
\textbf{First,} by selecting the most accurate model for in-cloud inference based on the network delay we increase accuracy within an overall latency target.
\textbf{Second,} by duplicating inference execution on-device using a low-latency model we can ensure that we can meet the latency target regardless of network connectivity and delay.
In short, by dynamically selecting a model while running inference both in-cloud and on-device we improve accuracy while providing latency guarantees for mobile applications.

 \begin{figure*}[t]
    \centering
    \includegraphics[width=0.9\textwidth]{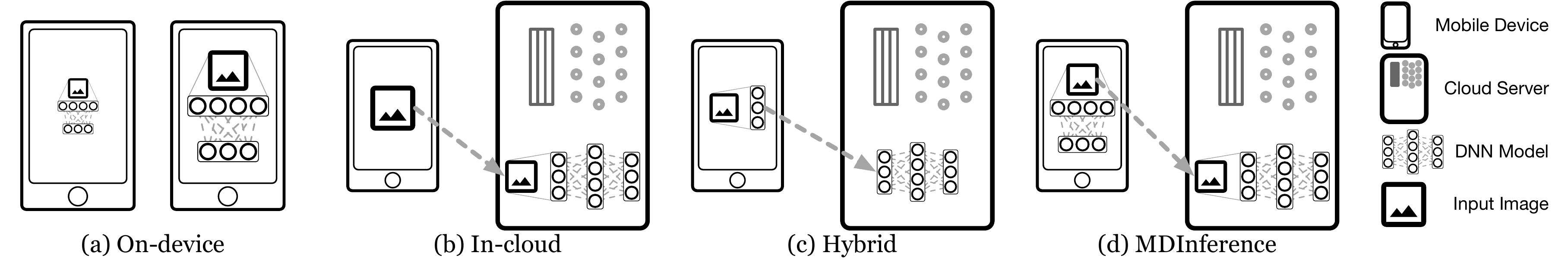}
    \caption{Comparison of different mobile inference techniques. \textnormal{\textit{(a) On-device inference allows models to run on-device in a resource constrained environment. Mobile-optimized models have lower latency but also lower accuracy. (b) Cloud-based inference allows for complex models but requires network transfer prior to inference execution, adding unpredictable network delay. (c) Hybrid inference shares execution between the mobile device and a cloud server, relying on both being available, to decrease latency. (d) \sysname uses runtime model selection and request duplication to select accurate cloud-based models and an on-device model to guarantee latency.}}}
    \label{fig:background:inference-approaches}
\end{figure*}

Our three main contributions are:
\begin{itemize}

\item {
    We introduce a new mobile-oriented approach to designing deep inference
    \frameworks that focuses on the specific goals and constraints of mobile
    devices, such as network condition variation.
    Making \frameworks aware of these constraints will allow them to improve \accuracy without sacrificing latency.
}

\item{
    We designed a hypothetical \framework \sysname that demonstrates the ability of this mobile-oriented approach to improve the \accuracy of inferences while meeting latency targets.
    Our evaluation shows \sysname achieved its target latency in 23\% more cases than in-cloud approaches and increased \accuracy over 39\% compared to purely on-device approaches.
}

\item{
    We developed and integrated two algorithms to enable our \sysname to be mobile-aware.
    These algorithms opportunistically increase the \accuracy of inferences and ensure that there are no SLA violations, improving user experience.
}

\end{itemize}

The remainder of this paper is structured as follows.
In Section~\ref{sec:background} we introduce a number of existing approaches mobile inference \frameworks. 
The problem of mobile deep inference is formalized in Section~\ref{sec:problem-statement}.
Section~\ref{sec:design} discusses the key advantages of existing approaches and describes how we design a hypothetical \framework for which we call \sysname.
An evaluation of the techniques implemented in \sysname is presented in Section~\ref{sec:eval} and a discussion of future directions is conducted in Section~\ref{sec:discussion}.

\section{Background and Motivation}
\label{sec:background}

Deep neural networks have become increasingly popular for embedding novel features into mobile applications.
Two common forms of deep learning, convolutional neural networks (CNNs) and recurrent neural networks (RNNs) excel at image processing and speech-to-text, respectively.
This allows for the addition of features such as Optical Character Recognition (OCR)~\cite{johnson2016googles} and virtual assistants~\cite{capes2017siri, kepuska2018next} to mobile applications.

State-of-the-art DNNs, with their accuracy-driven design, can contain tens of millions of parameters and hundreds of layers, and are therefore both computationally- and memory-intensive~\cite{He2015, Zoph2018, Szegedy2016, Bianco2018, guo2018cloud}.
To leverage these deep learning models, devices first need to preprocess the input data and load these models into memory.
Once these models are loaded into memory, executing them requires large matrix
multiplication operations with many millions, and often billions, of floating
point operations~\cite{Howard2017, Zoph2018, Szegedy2016}.
Therefore, while these models can add rich functionality to mobile applications, actually leveraging them on mobile devices is difficult due to resource constraints~\cite{guo2018cloud}.

Further, the number and extremity of otherwise common issues that mobile devices need to balance is extremely high.
First, mobile devices experience a wide range of network conditions both in terms of connection quality and speed.
They can be without a network connection for days or switch between high-speed WiFi and a cellular connection within the same minute.
Second, they are inherently resource constrained as devices must be small and efficient enough to be carried by end-users throughout daily life.
Finally, mobile applications are inevitably user-facing, compelling them to adhere to strict latency goals to improve user experience.

Three main approaches to enabling mobile deep learning are depicted in Figure~\ref{fig:background:inference-approaches} and discussed below.

\subsection{On-device Inference}
\label{subsec:background:on-device-inference}

\begin{figure}[t]
    \begin{center}
        \includegraphics[width=0.45\textwidth]{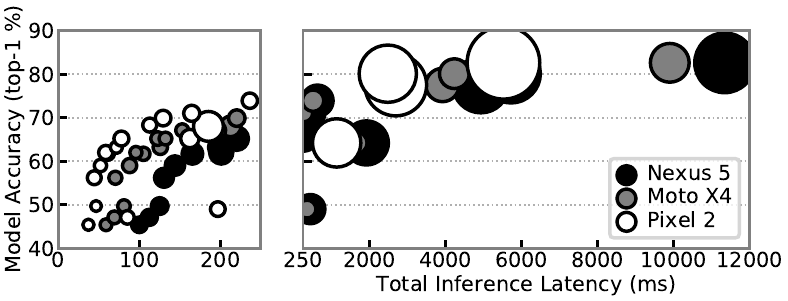}
        \caption{DNN execution latency on a range of mobile devices. \textnormal{\textit{ We measure execution time for 21 pretrained models~\cite{tensorflow:model-zoo} via the TensorFlow Lite framework~\cite{tensorflow:lite} running on three devices. The circle size corresponds to the standard deviation of the inference latency. We observe that high-accuracy models take multiple seconds to run on all devices and that newer devices, such as the Pixel 2, can support more models than older devices.}}}
        \label{fig:background:on-device-latency}
    \end{center}
\end{figure}

On-device inference refers to running DNNs on mobile devices, which is illustrated in Figure~\ref{fig:background:inference-approaches}(a), and is supposed by frameworks such as Caffe2~\cite{caffe2} and TensorFlow Lite~\cite{tensorflow:lite}.
These frameworks often use models that have been trained on powerful servers and exported to a format that is optimized for mobile devices.
Mobile oriented optimizations to decrease the latency of on-device execution often aim to reduce the complexity of the models themselves~\cite{Howard2017, Zoph2018}.

In Figure~\ref{fig:background:on-device-latency} we show the execution latency of 21 pretrained CNN models~\cite{tensorflow:model-zoo}.
While many of the models that have been optimized for mobile devices completed execution in under 250ms, these models have lower accuracy results.
Higher accuracy models often take much longer to run, even on devices with specialized hardware such as the Pixel 2~\cite{pixel2_specs}.
Further, we observe that the lower accuracy models show a distinct range of latencies, with latency increasing exponentially with accuracy, leading to the highest accuracy models having multi-second latency on all tested devices.

Further, even mobile-oriented models can still be orders of magnitude slower than running on dedicated servers with accelerators. 
The inference latency can be exacerbated when an application needs to load multiple models, such as chaining the execution of an OCR model and a text translation model~\cite{johnson2016googles}, or requires higher accuracy.

In summary, even though on-device inference is a plausible choice for simple tasks and newer mobile devices, it is less suitable for complex tasks or older mobile devices.

\subsection{In-cloud Inference}
\label{subsec:background:in-cloud-inference}

In-cloud inference, as illustrated in Figure~\ref{fig:background:inference-approaches}(b), executes models on remote servers instead of on-device.
Cloud-based servers, especially those with access to powerful accelerators such as GPUs, can execute models orders of magnitude faster than mobile devices.
For example, execution of the \emph{NasNet Large} model takes over 5 seconds on all of our mobile devices but takes only 113ms on a server with GPU (details in Table~\ref{eval:model-summary}).
By leveraging this decrease in latency, in-cloud inference could decrease overall response time, even while using more accurate models.
However, transferring the input data to the cloud-based servers can incur long and unpredictable network time~\cite{Ogden2018, Netravali2015}.

Model serving systems~\cite{Crankshaw2017, Olston2017, Gao:2018:LLR:3190508.3190541, LeMay2020} allow mobile applications to leverage these cloud-based \frameworks, often through REST APIs.
However, many such systems require mobile developers to \emph{manually} specify the exact model to use through the exposed API endpoints.
These frameworks fail to consider the impact of dynamic mobile network conditions, which can take up a significant portion of end-to-end inference time~\cite{Ogden2018, Satyanarayanan:2009:CVC:1638591.1638731}.
Moreover, such static development-time decisions can lead to using high-accuracy models whose high execution latency may be compounded by unexpectedly long network transfer time.

In summary, cloud-based inference has the potential to support many application scenarios, simple and complex, for heterogeneous mobile devices.
However, current mobile-agnostic serving platforms fall short by not automatically adapting inference choices based on mobile constraints.

\subsection{Hybrid Inference}
\label{subsec:background:hybrid-inference}

Hybrid inference spreads the execution of models across both the mobile device and a cloud-based server, as shown in Figure~\ref{fig:background:inference-approaches}(c).
By splitting the execution between two locations hybrid inference allows for decisions to be made at runtime to reduce overall response latency.

The division of model execution between the mobile device and the remote server is done by identifying partition points in models where intermediate data can be efficiently transferred from the mobile device to the remote server~\cite{Kang2017}.
Executing the first layers of a model on-device and then the rest of the model on a remote server allows for transferring less data across the network. 
However, if the network is unavailable the entire model can be executed locally, but an unpredictable network can lead to an increase in latency.

To remove this reliance on the network, each segment of the model execution can calculate a confidence in its response~\cite{Teerapittayanon2016, Teerapittayanon2017} where a high confidence will result in using the current response.
If the confidence is too low on-device, the intermediate data can be sent for remote inference.
This decreases the reliance on the network but potentially decreases accuracy.

In addition, since hybrid inference relies on continuing execution on the remote server this server has to host to same model as was used on the mobile device.
In order to accommodate the possibility of no network connection this limits the models that can be used for hybrid execution.

In summary, hybrid inference allows for decreasing latency by partitioning the inference model and selecting where and whether each of the pieces should be executed.
Network constraints may lead to longer latency and with a limited ability to improve accuracy by the models used. 

\section{Problem Statement}
\label{sec:problem-statement}

\begin{table}[t]
    \centering
    \resizebox{0.38\textwidth}{!}{
        \begin{tabular}{ll}
\multicolumn{1}{l|}{\textbf{Symbol}}                & \textbf{Meaning}                                                                                                       \\ \hline
\multicolumn{1}{l|}{$T_{sla}$}             & Response time SLA                                                                                             \\ 
\multicolumn{1}{l|}{$T_{budget}$}          & Time allowed for model execution                                                                              \\
\multicolumn{1}{l|}{$T_{nw}$}              & Estimated round-trip network time                                                                          \\
\multicolumn{1}{l|}{$\bm{M}$}              & A set of available models                                                                                       \\
\multicolumn{1}{l|}{$\bm{A}(m)$}           & Accuracy of a model $m$                                                                                        \\
\multicolumn{1}{l|}{$\mu(m)$, $\sigma(m)$} & \begin{tabular}[c]{@{}l@{}}Average and standard deviation of \\ model execution time for model $m$\end{tabular} \\
                                           &                                                                                                              
\end{tabular}
    }
    \caption{Symbols used throughout this paper.}
    
    \label{tbl:symbols}
\end{table}

We target the problem of designing mobile deep inference \frameworks for mobile applications. 
The core aspect of this problem is that the mobile device can have a variable, or nonexistent, network connection while request inferences.
Additionally, an application developer has access to a set of models $\bm{M}$ that exhibit a range of different accuracies and latencies~\cite{tensorflow:model-zoo, caffe:model-zoo} for the same task.
Therefore, the problem is about how to enable high-accuracy inference results for mobile devices within a given target latency.
Concretely, for a mobile device requesting an inference within a target latency, $T_{sla}$, we want to select an inference model, $m \in \bm{M}$ that maximizes accuracy and returns results within $T_{sla}$.
Note, all symbols used throughout this paper can be referenced in Table~\ref{tbl:symbols}.

We consider two main metrics that measure the quality of solutions to this problem.
\emph{First} is \textit{Service Level Agreement} (SLA) attainment, which is measured as the number of requests that return results within the specified response time target.
The goal for a mobile-oriented \framework is to return all results within a given SLA.
\emph{Second} is \emph{\accuracy}, which is the average accuracy of all models used by the \framework.
For example, if three inference requests are serviced by models with 40\%, 60\% and 60\% accuracy, then the \framework's \accuracy is 53.3\%.
The goal of any framework is to maximize its \accuracy.

\para{System model and assumptions:}
We assume our mobile device is resource constrained and can only run a single on-device model.
Further, we assume the mobile device \textit{may} have access to an in-cloud server hosting a set of functionally-equivalent models, but transferring input data can take a variable network time $T_{nw}$.
We call a set of distinct models \textit{functionally-equivalent} if they all perform the same task, such as image classification.
We further assume this network time can be calculated or estimated through a number of methods such as time synchronization, direct measurement, or network modeling~\cite{Netravali2015}.

Our hypothetical system is designed specifically for CNNs performing image classification tasks.
We assume that any required preprocessing is completed on the mobile device and is not directly considered as part of the response time.
We also assume that each request has an appropriate $T_{sla}$, representing the target request-response latency.

\section{Mobile Inference Framework Designs}
\label{sec:design}

Mobile-oriented inference \frameworks have a number of unique goals and constraints that we discuss next.
These represent a number of opportunities we discuss in Section~\ref{subsec:design:opportunities}.

\subsection{Mobile-Aware Framework Design Goals and Challenges}
\label{subsec:goals}

As more mobile applications are leveraging DNNs it is becoming critical that inference \frameworks be aware of the special demands of these applications.
Existing approaches focus on optimizing for single goals, such as latency on mobile devices or inference server throughput, while ignoring mobile-specific needs.
As an example, the \emph{NasNet Mobile} model was designed to provide high-accuracy inference on mobile devices.
On a Pixel 2 phone this model ran in 236ms but on other tested devices this model took up to 2.5X longer. 

\para{Goals for a mobile-specific inference framework:}
A mobile inference \framework needs to \emph{dynamically balance} two design goals: latency and accuracy.
This need is driven by a dynamic mobile environments and network connections, and the inherent heterogeneity of devices.

Latency is the time required to return an inference response to the mobile end-user.
Keeping this metric low and consistent is important to mobile applications which are inherently user facing.
Response times that are particularly long relative to the average will be obvious to users.

Accuracy is the ability a model to return the correct response on input data, which is often reported for image classifications models as the top-1 accuracy.
This describes the model's average likelihood to correctly classify input images.
In complex use cases accuracy is especially important.

\begin{table}[t]
    \resizebox{\columnwidth}{!}{
\begin{tabular}{l|c|c|c|c|c}
                   & \multicolumn{2}{c|}{\textbf{Goals}}                                            & \multicolumn{3}{c}{\textbf{Factors (Awareness)}}                                                                 \\ \cline{2-6} 
                   & \multicolumn{1}{l|}{\textbf{Accuracy}} & \multicolumn{1}{l|}{\textbf{Latency}} & \multicolumn{1}{l|}{\textbf{Network}} & \multicolumn{1}{l|}{\textbf{Resource}} & \multicolumn{1}{l}{\textbf{SLA}} \\ \hline
\textbf{On-Device} & \xmark                                 & \cmark                                & --                                    & \cmark                                 & \cmark                           \\ \hline
\textbf{In-Cloud}  & \cmark                                 & \xmark                                & \xmark                                & --                                     & \cmark                           \\ \hline
\textbf{Hybrid}    & \cmark/\xmark                          & \cmark/\xmark                         & \cmark                                & \cmark                                 & --                               \\ \hline
\textbf{\sysname}  & \cmark                                 & \cmark                                & \cmark                                & \cmark                                 & \cmark                           \\ \hline
\end{tabular}
    }
    \caption{Different mobile inference approaches and their goals and awareness. \textnormal{\textit{The three approaches discussed each have different optimization goals. On-device inference relies on an awareness of available resources to optimize for inference latency. In-cloud inference has the goal to increase the throughput of inference servers for the most accurate models, showing an attention to SLA but ignoring the network. With hybrid approaches, the goals and awareness lie on a spectrum. Typically \frameworks are aware of a subset of the various factors but no single approach is aware of all three. \sysname is aware of all three factors to achieve a reliable latency while increasing accuracy when possible.}}}
    \label{table:background:approach-summary}
\end{table}

\para{Challenges for mobile-oriented frameworks:}
An ideal mobile inference \framework should allow for both goals to be optimized by balancing them.
To do this it would have to be aware of three major constraints, which we introduce below and have summarized in Table~\ref{table:background:approach-summary}.

\emph{First}, mobile devices experience a wide range of network conditions that can lead to large variations in the latency of transferring input data for remote inference.
Frameworks that performs remote inference should be aware of this variation and able to adapt its inference decisions to minimize the impact.
\emph{Second}, mobile devices are inherently resource constrained, making on-device inference difficult, which is exacerbated by device heterogeneity.
A mobile-aware inference \framework should reduce its reliance on on-device inference as these constraints are device-specific and may force each device to use a different low-accuracy model.
\emph{Finally,} mobile applications are user facing and thus are generally very sensitive to response time.
Therefore any \framework providing mobile devices with inference services should be able to provide results within a reasonable time, often defined by its latency SLA.

\subsection{Inference Serving Opportunities}
\label{subsec:design:opportunities}

The existing approaches that mobile deep inference frameworks take  introduce a number of potentially opportunities.

\para{On-device inference aims to ensure that mobile users can always run inference but at a decreased accuracy.}
By decreasing the complexity of deep learning models it is possible to run inference directly on the mobile device within a reasonable latency.
This ensures that regardless of network connectivity mobile users can obtain inference results.
One example of this is \emph{MobileNets}~\cite{Howard2017} which by tuning the number of parameters within the model prior to training allows for a smooth trade-off curve between latency and accuracy based on the same model architecture.

The main drawback of on-device inference is that decreased latency is achieved by sacrificing inference accuracy.
In the case of MobileNets, this can mean decreasing the top-1 accuracy by 29.6\% (comparing the accuracy of the fastest and most accurate variations~\cite{tensorflow:model-zoo}).
The problem of trading accuracy for latency is further compounded by the need to make such decisions prior to training.
In particular, doing so at development time means an application either relies on a single model across all devices or needs to select the optimal model per device, which is challenging given the wide range of devices and models.

\para{In-cloud inference allows for high-accuracy models to be run with low latency but neglects the needs of mobile applications.}
By leveraging hardware accelerators such as GPUs, cloud-based inference servers can greatly reduce the latency and improve the serving throughput even with complex high-accuracy models~\cite{Crankshaw2017, Olston2017, Zhang2019}.
As an example, we observed that the time to execute the \emph{NasNet Large} model (82.6\% accuracy) in the cloud was faster than running inference requests with the \emph{MobileNetV1\_160 1.0} model (68.0\% accuracy) on the fastest mobile device in our experiments.
(For details see Figure~\ref{fig:background:on-device-latency} and Table~\ref{eval:model-summary}.)
Cloud-based serving allows not only for high-accuracy inferences with low execution latency, but also opens up opportunities to serve inference requests with functionally-equivalent models that exhibit different latency-accuracy trade-offs.

The drawback of in-cloud inference \frameworks is that they mobile-agnostic and are typically oriented towards service providers.
This has two impacts.
First, cloud-based servers aim to achieve a service level objective considering only on-server time and exclude the network latency of the input data~\cite{Kannan2019, Crankshaw2017}.
Due to this, poor mobile network connections can result in poor mobile performance~\cite{Ogden2018}.
Second, optimizations for throughput, such as batching, lead to an increase in the latency of individual requests~\cite{Crankshaw2017, Bianco2018}.

\para{Hybrid inference spreads execution across multiple locations allowing for decreased latency but at the cost of relying on the availability of both locations.}
Spreading inference across multiple devices allows for a decrease in the amount of data transmitted across the network~\cite{Kang2017} or to exit early from execution when confidence in the intermediate result is above a threshold~\cite{Teerapittayanon2017}.
As a result, \frameworks that support hybrid inference have the flexibility to selectively improve the inference performance by carefully spreading the model across different locations.

However, this requires both that intermediate data be transferred between locations and that the intermediate data can be used in both locations, leading to the same model be executed in both locations.
In the case that network transfer of intermediate data is prohibitive the model must be executed entirely on-device.
For complex models this leads to unacceptable latency, and simple models fail to benefit from the remote execution.
Therefore, hybrid \frameworks have similar limitations to on-device \frameworks, in that the model used must be selected during development, and in-cloud \frameworks with their sensitivity to the network.

\section{\sysname Framework Design}
\label{subsec:design:system-design}

The key insight of \sysname is that we can leverage a set of cloud-based functionally-equivalent models to improve accuracy.
In addition, duplication of inference requests~\cite{Ousterhout2013, Mitzenmacher2001} allows us to bound latency.
For each inference \sysname submits an inference to a remote server that dynamically selects an accurate model, and at the same time executes a low model to ensure results will be available for uses within the SLA.
This allows for increased accuracy and reliable latency.

\sysname combines the advantages of existing approaches in order to improve end-user performance.
By dynamically selecting cloud-based models based on network information we can opportunistically use higher accuracy models and improve the \accuracy.
Additionally, \sysname and further improve the \accuracy by using a more accurate on-device model, although this can impact the minimum achievable SLA.
This combination of local and remote inferences allows \sysname to provide for reliable latency and improved \accuracy.

\sysname consists of two components.
First, a cloud-based server selects between a number of functionally-equivalent models for one that can complete within a specific SLA by estimating the time consumed for transferring input data.
This algorithm is detailed in Section~\ref{subsec:design:algorithm}.
Second, a local inference is run on-device to ensure that results are available within the target SLA.
The combination of these two components ensures that inference output is available within the SLA, possibly with improved \accuracy from the cloud-based component.
We discuss the implication of duplicating inference requests in Section~\ref{subsec:design:duplication}.

\subsection{Model Selection Algorithm}
\label{subsec:design:algorithm}

\sysname's model selection algorithm is designed to manage a set of functionally-equivalent CNN models and pick the most accurate model that can return results within the specified SLA.
It is designed to take advantage of the low variability of model execution latency to not only mitigate the impact of variations in the mobile network latency but opportunistically use them to improve accuracy.
The key insight of our model selection algorithm is that the variations of transfer latency for an inference request can be compensated for with the appropriate choice of inference model.
As functionally-equivalent models each have different execution times and accuracies, by explicitly making inference latency and model accuracy trade-offs \sysname determines which CNN model to execute.

\sysname works by selecting the most accurate model that has a low enough execution time to return results to the end-user within the SLA.
It accomplishes this by first calculating the request's time budget as the difference between SLA and the estimated network time.
That is, $T_{budget} = T_{sla} - T_{nw}$ where $T_{nw}$, referred to as \emph{network time}, denotes the time to transfer the inference request and to return the result.
Consequently, $T_{nw}$ can be estimated conservatively as $T_{nw} = 2 \times T_{input}$ where $T_{input}$ is the time to transfer the data to the remote server.
Estimating $T_{nw}$ using $T_{input}$ is straightforward as $T_{input}$ can be measured by the server prior to inference execution.
Further, such estimation is reasonable for application scenarios such as image recognition or image-to-text translations.
These applications often need to send more data to the cloud (i.e., input data) which leads to $T_{input} \geq T_{output}$, the time to return results. 
For other application scenarios such as speech recognition where output data size is often larger, one could leverage past observations of $T_{output}$ and estimate $T_{nw} = 2 \times T_{output}$ instead.
Using this time budget we can then identify the set of models, $M_{E}$, that can complete execution within the request time budget $T_{budget}$.

The basic approach described above assumes that the execution times and accuracies of models previously measured stays the same.
However, these two assumptions do not always hold, leading to a need to expand on the basic concept of model selection  to \textit{probabilistically} select models.
Real-world serving systems~\cite{Olston2017, Crankshaw2017} often experience queuing delay or workload spikes~\cite{Bodik2010} leading to transient increases in latency.
Additionally, accuracy is affected over time by concept drift~\cite{Widmer1996}.
To handle these changes in latency and accuracy the model selection algorithm probabilistically selects models, thus exploring available models that might have been previously disregarded due to transient issues.
We do this by selecting a model using a weighted probability based on the model's latency relative to $T_{budget}$ and accuracy. 
We implement this probabilistic approach via a three stage algorithm described below.

\para{Stage one: greedily picking the baseline model.}
In this stage, \sysname takes all the existing models and selects a base model $m_j$ as follows.
\begin{alignat}{4}
& \underset{j}{\text{maximize}} & \hspace{5mm}
&  \bm{A} (m)  \label{eq:obj}\\
& \text{subject to} && \mu(m) + \sigma(m) < T_{budget} , \; m \in \bm{M}
\label{eq:c1}
\end{alignat}

To find the base model we first consider all models that have an expected inference time less than the time budget and use the most accurate of these models.
This is to make it likely that the model will finish within the calculated budget.
We use this model $m_b$ as our base model.
In the case that no models satisfy the time budget constraint the fastest model available is chosen as the base model and execution begins.

\para{Stage two: optimistically constructing the eligible model set.} 
In order to account for unexpected performance variations, such as queueing delays or accuracy variations, the probabilistic model selection algorithm will expand around the base model to form an exploration set, $M_E$.
This exploration set represents models that are similar to the base model in terms of execution time.
Specifically, we construct the exploration set as 
\begin{equation}
M_E = \{m \mid \mu(m) \in [\mu(m_b) - \sigma(m_b), \mu(m_b) + \sigma(m_b)] \}
\end{equation}
 which is the set of all models that have an average execution time within the typical execution time of the base model.
It is important to note that $M_E$ may include models that violate the latency variation constraints imposed on the base model.
This is accounted for in stage three.

\para{Stage three: opportunistically selecting the inference model.} 
From the exploration set $M_E$ we select a model $m'$ to balance the risk of SLA violations and the exploration reward.
Concretely, we calculate the utility for each model, $\bm{U}(m)$, based on its inference accuracy and its likelihood to violate response time SLA as:

\begin{equation}
\bm{U}(m) = \bm{A}(m) \frac{T_{budget} - \big(\mu(m) + \sigma(m)\big)}{| T_{budget} - \mu(m) |} .\label{eq:utility}
\end{equation}

\sysname than normalizes these utilities to calculate the selection probability as
$\bm{Pr}(m) = \frac{\bm{U}(m)}{\sum\limits_{n \in \bm{M_E}} \bm{U}(n)}$ 
and picks $m'$ based on its probability.
This helps avoid choosing models with lower inference accuracy, wider inference time distribution, and outdated performance profile while still exploring the set of potentially eligible models.

\subsection{Request Duplication}
\label{subsec:design:duplication}

To ensure that all requests can be serviced within the SLA, \sysname duplicates requests to bound their tail response latency.
As discussed in Section~\ref{subsec:background:on-device-inference}, many mobile-oriented models can produce results on-device within a reasonable time limit, but with lower accuracy.

When an inference is initiated two requests are generated by the \sysname \framework.
The first is sent to a remote inference server that executes the model selection algorithm outlined previously.
While this cloud request aims to return results within the SLA it is not guaranteed.
Therefore, an inference request is duplicated and executed locally using the on-device model.
In \sysname we chose the fastest available model to use on-device, supporting for SLAs as low as 50ms, although any model that satisfies the SLA goal can be used.

There are two potential outcomes to duplication.
First, the SLA expires without the remote inference request having returned results, in which case \sysname uses the results of the on-device model.
In our experiments this occurred in only 3.16\% of cases, as we will see in Section~\ref{subsec:eval:duplication}.
The second outcome has the remote inference response arrives before the SLA expires and the remote results are used.

\begin{table}[t]
    \resizebox{\columnwidth}{!}{
    \begin{tabular}{|l|l|l|l|}
    \hline
    \textbf{Model Name}     & \textbf{Top-1 Accuracy (\%)} & \textbf{Inference  Avg. (ms)} & \textbf{Inference Std. (ms)} \\ \hline
    \textbf{SqueezeNet}              & 49.0                    & 4.91 &  0.06        \\ \hline
    \textbf{MobileNetV1 0.25}        & 49.7                    & 3.21 &  0.08      \\ \hline
    \textbf{MobileNetV1 0.5}         & 63.2                    & 4.21 &  0.06         \\ \hline
    \textbf{DenseNet}                & 64.2                    & 25.49 &  0.14        \\ \hline
    \textbf{MobileNetV1 0.75}        & 68.3                    & 4.67 &  0.07         \\ \hline
    \textbf{MobileNetV1 1.0}         & 71.0                    & 5.43 &  0.11         \\ \hline
    \textbf{NasNet Mobile}           & 73.9                    & 21.18 &  0.17        \\ \hline
    \textbf{InceptionResNetV2}       & 77.5                    & 50.85 &  0.33        \\ \hline
    \textbf{InceptionV3}             & 77.9                    & 31.11 &  0.19       \\ \hline
    \textbf{InceptionV4}             & 80.1                    & 59.21 &  0.22        \\ \hline
    \textbf{NasNet Large}            & 82.6                    & 112.61 & 0.36       \\ \hline
    \textbf{NasNet Fictional*}   & 50 & 112.61 & 0.36       \\ \hline
    \end{tabular}
    }
    \caption{Summaries of model statistics through empirical measurement. \textnormal{\textit{Models are sorted based on their reported top-1 accuracy which is defined as the percentage of correctly labeled test images using only the most probable label. We measure the average inference time and standard deviation  for each model running via TensorFlow on an AWS \textit{p2.xlarge} GPU server. We used these models in simulations to study \sysname's effectiveness in trading-off \accuracy and latency. Note, \emph{NasNet Fictional} is a copy of \emph{NasNet} with lower accuracy used in Section~\ref{subsec:decomposing_benefits}.}}}
    \label{eval:model-summary}
\end{table}

\section{Evaluation}
\label{sec:eval}

Our evaluation goal is to quantify the effectiveness of \sysname in opportunistically improving \accuracy for mobile devices.
We do this by first demonstrating the effectiveness of our model selection algorithm at increasing the \accuracy at a range of SLAs when compared to a number of alternative algorithms.
Finally, we demonstrate the benefit of request duplication by analyzing its impact on SLA attainment and \accuracy.

\para{Key metrics.}
The first metric that we are measuring in many of our evaluations is \accuracy, which we introduced in Section~\ref{sec:problem-statement}.
We additionally measure the SLA attainment, which is the percent of requests that return results to the user within a target SLA.
With duplication this is no longer an issue thanks to leveraging low-latency on-device models.
In these cases we measure the percentage of requests that rely on the on-device model and the \accuracy improvements when leveraging in-cloud models.

\para{Simulation methodology.}
In our simulations, we leverage a range of models, summarized in Table~\ref{eval:model-summary}\cite{tensorflow:model-zoo, Howard2017, Szegedy2016, imagenet_cvpr09, Iandola2016, Simonyan2014, He2015}, that expose different accuracy and inference time trade-offs.
We empirically measured the inference time distributions of models using an EC2 \textit{p2.xlarge} GPU-accelerated server over 1,000 inference executions.
The accuracy of our functionally-equivalent pretrained model was reported against the ILSVRC 2012 dataset~\cite{ILSVRC15, tensorflow:model-zoo}, a widely used image classification test set.
The mobile network we use as the basis for many of our simulations assumes that transferring an input image takes $100$ms$\pm50$ms, based on real world measurements of our university network.
For each simulation, we generate 10,000 inference requests with a predefined SLA target and recorded the model selected by \sysname (and baseline algorithms) and relevant performance metrics.
We repeated each simulation for a variety of SLA targets and network profiles combinations.
For all tests except those in Section~\ref{subsec:eval:duplication} we evaluated the model selection capabilities of \sysname without duplication of requests.

\begin{figure}[t]
    \begin{subfigure}{0.5\textwidth}
    \centering
    \includegraphics[width=0.9\textwidth]{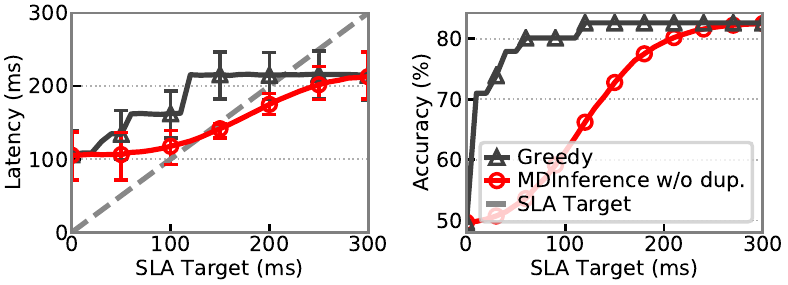}
        \caption{ \textnormal{\textit{Comparison of average end-to-end latency and accuracy. \sysname's selection algorithm is able to track the SLA target when SLA $\geq$100ms while the greedy approach fails to do so. \sysname improved \accuracy \emph{safely} as the SLA target increases. Further, our model selection algorithm improves standard deviation of inference times as well keeping them below the SLA target.}}}
    \label{eval:subfig:two-models:compare} 
    \end{subfigure}
    \hfill
    \begin{subfigure}{0.5\textwidth}
    \centering
    \includegraphics[width=0.9\textwidth]{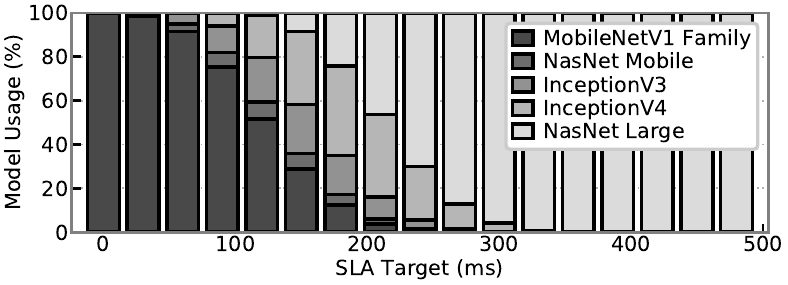}
        \caption{ \textnormal{\textit{Models selected by \sysname. The use of a diverse set of models allows \sysname to minimize SLA violations while providing highly accurate inference results.}}}
    \label{eval:subfig:two-models:model-usage} 
    \end{subfigure}
    \caption{Comparison of \sysname to the \emph{static greedy} algorithm. \textnormal{\textit{For each SLA target, we simulated 10,000 inference requests and recorded the inference time incurred by both the greedy algorithm and \sysname.}}}
    \label{eval:fig:two-models}
\end{figure}

\subsection{Benefits over static greedy model selection}
\label{sec:eval:benefits-of-multimodel-hosting}

Figure~\ref{eval:subfig:two-models:compare} shows the average end-to-end inference time (left) and \accuracy (right) achieved by our model selection algorithm and a \textit{static greedy} algorithm, which picks the most accurate model that can complete within the given SLA.
This figure shows that \sysname consistently achieved up to 42\% lower inference latency, compared to \emph{static greedy}.
Moreover, \sysname can operate under a much more stringent SLA target ($\mathtt{\sim}115ms$) while \emph{static greedy} continues to frequently incur SLA violations until SLA target is more than 200ms.
The key reason is that \sysname was able to effectively trade off \accuracy and response time by choosing from a diverse set of functionally-equivalent models.
Consequently, \sysname had an \accuracy of 68\% (on par to using \emph{MobileNetV1 0.75} which can take 2.9X more time running on mobile devices) under low SLA target ($\mathtt{\sim}115ms$), but was able to match the \accuracy achieved by \emph{static greedy} for higher SLA target.
Note that \emph{static greedy} achieved up to 12\% higher accuracy by sacrificing inference latency. 

Figure~\ref{eval:subfig:two-models:model-usage} illustrates the CNN model usage patterns (i.e., percentage of model being used for executing the inference requests) under different SLA targets.
At very low SLA target (less than 30ms), \sysname chooses the fastest model, \emph{MobileNetV1 0.25}, as described in Section~\ref{subsec:design:algorithm}.
As the SLA target increases, \sysname aggressively explores more accurate, but slower models, commonly using our most accurate model, \emph{NasNet Large}.

We make two observations.
First, \sysname was effective in picking the more appropriate model to increase accuracy while staying safely within SLA target.
For example, \emph{InceptionResNetV2} was never selected by \sysname because better alternatives such as \emph{InceptionV3} and \emph{InceptionV4} exist.
Second, \sysname faithfully explored eligible models and was able to converge to the most accurate model when SLA target allows, as shown in Figure~\ref{eval:subfig:two-models:compare} at $T_{sla} = 250ms$.

\textbf{In summary}, \sysname outperformed \emph{static greedy} with an end-to-end latency reduction of up to 43\%, while matching its accuracy when the SLA budget is larger than $250$ms.
This is possible because \sysname adapted its model selection by considering both the SLA target and network transfer time, while \emph{static greedy} naively selected the most accurate model.

\begin{figure}[t]
    \centering
    \includegraphics[width=0.45\textwidth]{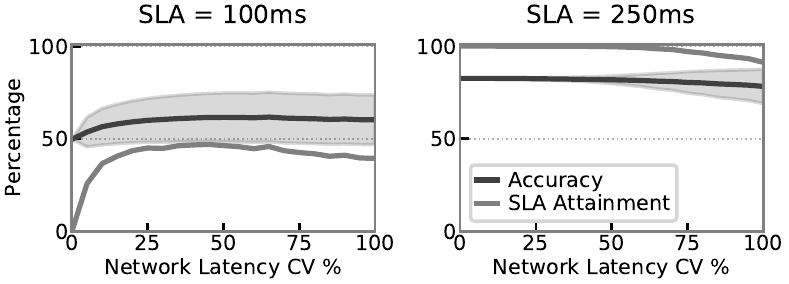}
        \caption{\Accuracy of \sysname at different levels of CV with $T_{nw}$ = 100ms. \textnormal{\textit{The initial low level of SLA attainment is due to the initial network time of 100ms, leaving no time for inference execution. As the variability of the network increases \sysname can take advantage of the range of models available to it to quickly improve accuracy and SLA attainment. Similarly, at a higher SLA, \sysname can achieve high accuracy until the network variability exceeds the SLA.}}}
    \label{eval:fig:CV:accuracy}   
\end{figure}

\begin{figure}[th]
    \centering
    \includegraphics[width=0.45\textwidth]{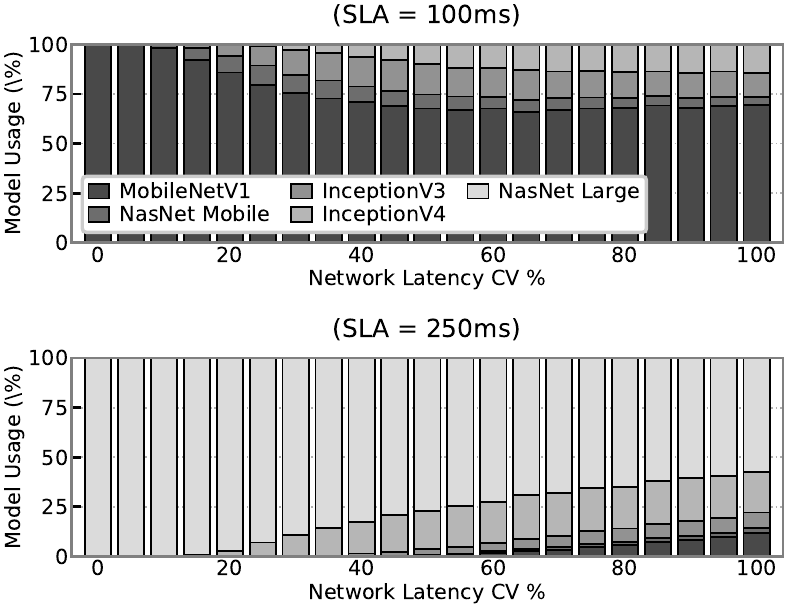}
    \caption{ Model usage vs. network latency (CV) shown at two different SLA targets. \textnormal{\textit{ When there is a reliable network (i.e. low CV) we see that single models are used for all requests since the high reliable network has leaves no room for other model choices. As the network becomes increasingly volatile a wider range of models is used to meet the SLA.}}}
    \label{eval:fig:CV:model-usage}
\end{figure}

\subsection{Adaptiveness to dynamic mobile network conditions}
\label{eval.subsection.dynamic_network_conditions}

One of the key goals of \sysname is to adapt to network variations in order to improve mobile user experience.
To further examine how \sysname copes with these variations we simulated network profiles with increasing variability.
Specifically, we fixed the average network latency to be $100$ms, and varied the Coefficient of Variation (CV) from 0\% to 100\%, where CV is defined as the ratio between the standard deviation and average of the network time.
CV ranged from 0\% to 100\%, to represent a perfectly stable network and a network which has a standard deviation equals its average, respectively.
As a point of reference, our measured university WiFi network has a CV of 74\%.

Figure~\ref{eval:fig:CV:accuracy} shows the \accuracy and SLA attainment achieved by \sysname.
For low SLA target (100ms), when the network is relatively stable \sysname had an SLA attainment of less than 50\%.
As the network condition became more variable, \sysname was able to increase the \accuracy gradually while maintaining the SLA attainment.
The low SLA attainment is due to on average half the requests needing the entire SLA just for network transfer.
Conversely, the slight increase in accuracy is due to \sysname taking advantage of the network variation to use more accurate models.

It is important to note that when the network transfer took the entirely SLA, \sysname performed as expected by choosing \emph{MobileNetV1 0.25}, the fastest available model.
We note that, to satisfy such stringent SLA targets with high network latency variation, approaches such as provisioning inference servers at network edge. 

When we consider an SLA target of 250ms we see a different outcome.
This SLA is slightly more than the sum of the average network latency and the time to execute the most accurate model, \emph{NasNet Large}.
In this case we see that \sysname used high accuracy models, maintaining an accuracy of around 80\% throughout the entire range of network variations.

Figure~\ref{eval:fig:CV:model-usage} shows the models chosen when varying CV of network time for SLAs of 100ms and 250ms.
The SLA of 100ms is the RTT of the simulated network leaving no time left for inference.
Similarly, when the SLA is 250ms then the target time is larger than the sum of the RTT of the network and the execution latency of \emph{NasNet Large}, our most complex model, which \sysname leverages.

We make the following two observations.
First as the network became more variable (i.e. high CV), \sysname matched the network variability by using a subset of faster models.
As the network becomes more variable \sysname can exploit this to in some cases opportunistically use models with high inference accuracy.
Second, the probability of exploring different eligible models is proportional to the SLA target and network variability.
Faster models, such as those in the \emph{MobileNetV1} family, are used as a basis for low SLA target while the most accurate model, \emph{NasNet Large}, is used for higher SLA targets.

\textbf{In summary}, \sysname was effective in handling highly variable mobile network by exploring a diverse set of deep learning models that expose different inference latency and accuracy trade-offs.
By taking advantage of network variation it could improve accuracy in many cases and even in cases with low target response times could often return responses within the SLA.

\subsection{Decomposing the efficiency of \sysname}
\label{subsec:decomposing_benefits}

\begin{figure}[t]
    \centering
    \includegraphics[width=0.45\textwidth]{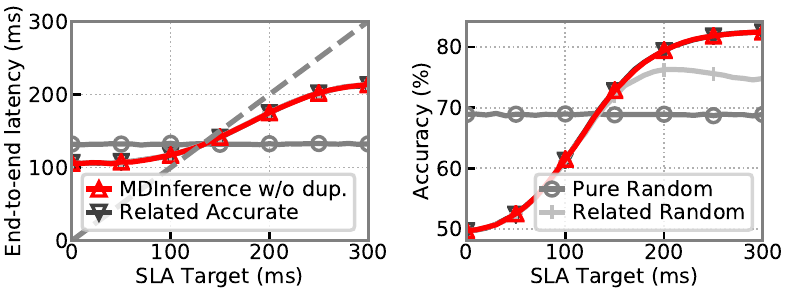}
        \caption{Decomposition of benefits of \sysname's three-stage algorithm. \textnormal{\textit{\sysname achieves similar accuracy and SLA attainment compared to \emph{related accurate}, indicating the effectiveness of our probabilistic approach. Both \emph{pure random} and \emph{related random} have poor \accuracy due to their inability to distinguish models with different latency.}}}
    \label{eval:fig:algos:compare} 
\end{figure}

Next, we breakdown the performance benefits provided by \sysname by examining the stages of our probabilistic model selection algorithm (Section~\ref{subsec:design:algorithm}).
For each of the three stages we compare to an alternative algorithm.
For stage one we compare to \emph{random} model selection.
For stage two we compare to \emph{related random} that randomly selects a model from the exploration set $M_E$.
For stage three we compare to \emph{related accurate}, which selects the most accurate model from $M_E$ to demonstrate that our probabilistic approach does not sacrifice accuracy.
All algorithms were tested with a network latency with average $100ms$ and standard deviation of $50ms$.

Figure~\ref{eval:fig:algos:compare} shows the average inference latency and \accuracy for all four model selection algorithms.
All three algorithms, including \sysname, that choose from the exploration set $M_E$ were able to meet reasonable SLA target while \emph{pure random} had approximately the same latency across all SLAs, incurring a large number of SLA violations.
This indicates that the construction of $M_E$, by stage one and stage two both enabled exploration and minimized risk of unnecessary SLA violations. 

As the SLA target increases, \emph{pure random} again achieved approximately the same \accuracy across all SLAs.
All three other algorithms were able to gradually increase the \accuracy by using slower but more accurate models from Table~\ref{eval:model-summary}.
However, once we have a large enough SLA target ($\mathtt{\sim}150$ms), the exploration set $M_E$ converges to two models: \emph{NasNet Large} and \emph{NasNet Fictional}.
At this point, \emph{related random} algorithm started to degrade \accuracy since it cannot differentiate between these two models.  
Meanwhile, both \emph{related accurate} and \sysname were able to steadily improve \accuracy by avoiding \emph{NasNet Fictional}.

It is important to note there is only a negligible difference in \accuracy using \sysname when compared to \emph{related accurate} algorithm.
This small difference is due to \emph{related accurate} always selecting the most accurate model from $M_E$ while \sysname explores the eligible set. 
The probability of picking a less accurate model is low enough as to not overly impact the \accuracy.
The probabilistic behavior of \sysname is meant to allow for this exploration even while generally maintaining \accuracy, as opposed to \emph{related accurate}, which misses the opportunity to use models which may have improved accuracy or latency profiles.

\textbf{In summary}, \sysname's three-stage algorithm is effective in distinguishing and identifying the most appropriate model to use under dynamic conditions.
All three stages contribute to and help \sysname cope with the variable network conditions and improve \accuracy.

\subsection{Effectiveness of Request Duplication}
\label{subsec:eval:duplication}

\begin{table}[t]
    \resizebox{\columnwidth}{!}{
\begin{tabular}{l|c|c|c|r}
                         & \multicolumn{2}{c|}{\textbf{University}}                                                                                                                                                & \multicolumn{2}{c|}{\textbf{Residential}}                                                                                                                                               \\ \cline{2-5} 
                         & \multicolumn{1}{l|}{\textbf{\begin{tabular}[c]{@{}l@{}}On-device\\ Reliance\end{tabular}}} & \multicolumn{1}{l|}{\textbf{\begin{tabular}[c]{@{}l@{}}Aggregate\\ Accuracy\end{tabular}}} & \multicolumn{1}{l|}{\textbf{\begin{tabular}[c]{@{}l@{}}On-device\\ Reliance\end{tabular}}} & \multicolumn{1}{l|}{\textbf{\begin{tabular}[c]{@{}l@{}}Aggregate\\ Accuracy\end{tabular}}} \\ \hline
\textbf{Static Latency}  & \cellcolor[HTML]{9AFF99}0.26\%                                                             & 41.40\%                                                                                    & \cellcolor[HTML]{9AFF99}3.16\%                                                             & \multicolumn{1}{r|}{41.40\%}                                                               \\ \hline
\textbf{Static Accuracy} & 3.67\%                                                                                     & 81.09\%                                                                                    & 23.03\%                                                                                    & \multicolumn{1}{r|}{73.11\%}                                                               \\ \hline
\textbf{Random}          & 0.42\%                                                                                     & 63.33\%                                                                                    & 5.06\%                                                                                     & \multicolumn{1}{r|}{62.06\%}                                                               \\ \hline
\textbf{\sysname}        & \cellcolor[HTML]{9AFF99}0.26\%                                                             & \cellcolor[HTML]{9AFF99}82.39\%                                                            & \cellcolor[HTML]{9AFF99}3.16\%                                                             & \multicolumn{1}{r|}{\cellcolor[HTML]{9AFF99}80.43\%}                                       \\ \hline
\end{tabular}
    }
    \caption{\Accuracy and on-device model reliance. \textnormal{\textit{\sysname achieved the highest \accuracy with lower on-device reliance, compared to other algorithms.}}
    }
    \label{table:eval:on-device-usage}
\end{table}

\begin{figure}[t]
    \centering
        \includegraphics[width=0.45\textwidth]{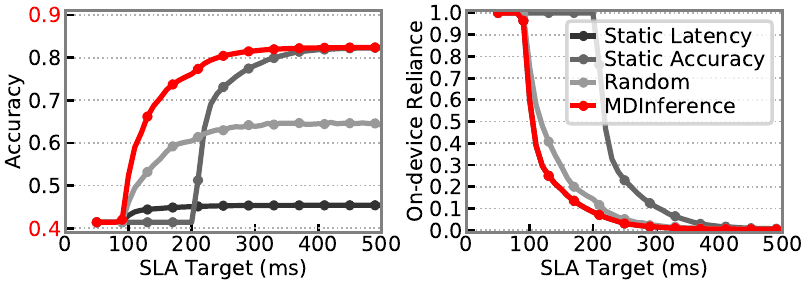}
        \caption{\Accuracy and on-device model reliance on residential network.
        \textnormal{\textit{\sysname demonstrated improvements throughout all tested SLAs. At lower SLAs \sysname can quickly improve the \accuracy. Meanwhile, \sysname also reduces the reliance on on-device models at low SLAs, much more quickly than other algorithms.}}}
    \label{fig:eval:on-device-usage} 
\end{figure}

\begin{figure}[t]
    \centering
        \includegraphics[width=0.45\textwidth]{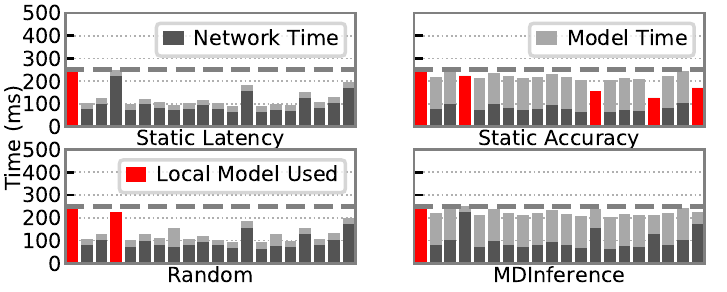}
        \caption{Inference time for 20 randomly sampled requests over
        residential network. \textnormal{\textit{In many cases \sysname chose a model that yields results within the time budget. Other approaches, such as Static Accuracy, returned high-accuracy results but rely more heavily on the on-device model due to network variation.}
        }}
    \label{fig:eval:on-device-usage-specific-requests} 
\end{figure}
\vspace{-5mm}

To test the on-device reliance of \sysname we used the network time from sample of 5000 requests on each of our university network and on a residential network.
These requests consisted of a preprocessed image input that averaged 51.9KB with standard deviation of 53.6KB.
The model chosen for the on-device was the fastest \emph{MobileNetV1\_128 0.25} model as it represents the single model most likely to complete within any SLA for all tested mobile devices.
It was also excluded from the set of models available in the cloud to better demonstrate the ability of \sysname to improve over on-device inference.

For each of these measured requests we compared \sysname to three other simulated model selection algorithms using the models detailed in Table~\ref{eval:model-summary} and an SLA target of 250ms.
The three other algorithms we used were \emph{static latency}, which picks the fastest model, \emph{static accuracy}, which picks the most accurate model, and \emph{random} which picks a random model.

Table~\ref{table:eval:on-device-usage} compares the \accuracy and on-device reliance for all four model selection algorithms.
\sysname achieved the highest \accuracy for inference requests sent over both university and residential WiFi, improving over static accuracy by 7.32\% on the variable residential network.
Meanwhile, it improved \accuracy compared to static accuracy by up to 19\%.

The \accuracy and on-device reliance is shown in Figure~\ref{fig:eval:on-device-usage}.
We can observe that \sysname increases \accuracy more quickly than the other algorithms tested and has a lower reliance on the on-device models, allowing it to maintain this higher \accuracy.

Figure~\ref{fig:eval:on-device-usage-specific-requests} shows the inference latency breakdown for 20 randomly sampled requests that were sent over the residential network.
We observe that mosts requests were able to complete on the remote server but in some cases the on-device model must be used, in which we highlight the network latency in red.
This shows the ability of \sysname to adapt its execution choice to match compensate for network variability, allowing it to decrease its on-device reliance and boosting is accuracy.

\textbf{In summary,} the duplication mechanism allows \sysname to ensure that results are returned to the mobile user within the target response time.
Combining this with the model selection algorithm, \sysname is able to increase the \accuracy for inference requests in the vast majority of cases.

\section{Discussion \& Future Work}
\label{sec:discussion}

There are a number of potential avenues for future work in mobile-oriented deep inference \frameworks.
We discuss a number of important factors that should be considered, such as energy consumption and \accuracy.

\para{Energy Consumption.}
The duplication of inference requests solves the issues of SLA violations, allowing users to have reasonable performance.
However, this requires energy consumption on the device for both network communication and inference.
Therefore, identifying times when duplication is critical and avoiding unnecessary duplications could allow for reduced energy consumption.

\para{On-device Model Selection.}
Currently, our proposed \sysname \framework uses the same on-device model regardless of the mobile devices.
There are a number of different approaches, discussed in Section~\ref{subsec:background:on-device-inference} for improving on-device inference but generally rely on statically selected models.
While some model optimizations can provide this ability to simplify models post-training~\cite{Han2015}, these provide only limited options.

\para{Spanning Subsets of Models.}
Figure~\ref{eval:subfig:two-models:model-usage} demonstrated that many requests can be serviced by only a small subset of models.
This potentially indicates that there exists a subset of models that could service nearly all requests, and thus form a \emph{spanning subset} for all the models.
This would be highly beneficial for decreasing the cost of inference serving, as only the models that fall into this subset would need to be available.
Further, finding this subset for an arbitrary set of models without resorting to empirical measurement is another challenging problem to investigate.

\section{Conclusion}
\label{sec:conclusion}

In this work we introduced a holistic approach to designing mobile-oriented deep inference \frameworks that focuses on identifying user needs and the constraints of mobile devices.
We introduced the design of \sysname, a hypothetical framework utilizing this approach.
\sysname can improve \accuracy in over 96\% of cases without introducing additional SLA violations.
This improvement in accuracy was over 40\% in some cases and was a 7.32\%
increase over statically serving the highest-latency model while duplicating
inference execution locally. 
Our work shows the potential to better mobile inference serving by explicitly addressing mobile-oriented constraints.

\section*{Acknowledgment}
We would like to thank our anonymous reviewers for their valuable feedback.
This work was supported in part by NSF Grants \#1755659 and \#1815619.

\balance

\bibliographystyle{IEEEtran}
\bibliography{../bib/bib}

\begin{thebibliography}{10}
\providecommand{\url}[1]{#1}
\csname url@samestyle\endcsname
\providecommand{\newblock}{\relax}
\providecommand{\bibinfo}[2]{#2}
\providecommand{\BIBentrySTDinterwordspacing}{\spaceskip=0pt\relax}
\providecommand{\BIBentryALTinterwordstretchfactor}{4}
\providecommand{\BIBentryALTinterwordspacing}{\spaceskip=\fontdimen2\font plus
\BIBentryALTinterwordstretchfactor\fontdimen3\font minus
  \fontdimen4\font\relax}
\providecommand{\BIBforeignlanguage}[2]{{%
\expandafter\ifx\csname l@#1\endcsname\relax
\typeout{** WARNING: IEEEtran.bst: No hyphenation pattern has been}%
\typeout{** loaded for the language `#1'. Using the pattern for}%
\typeout{** the default language instead.}%
\else
\language=\csname l@#1\endcsname
\fi
#2}}
\providecommand{\BIBdecl}{\relax}
\BIBdecl

\bibitem{capes2017siri}
T.~Capes, P.~Coles, A.~Conkie, L.~Golipour, A.~Hadjitarkhani, Q.~Hu,
  N.~Huddleston, M.~Hunt, J.~Li, M.~Neeracher \emph{et~al.}, ``Siri on-device
  deep learning-guided unit selection text-to-speech system.'' in
  \emph{INTERSPEECH}, 2017, pp. 4011--4015.

\bibitem{kepuska2018next}
V.~Kepuska and G.~Bohouta, ``Next-generation of virtual personal assistants
  (microsoft cortana, apple siri, amazon alexa and google home),'' in
  \emph{2018 IEEE 8th Annual Computing and Communication Workshop and
  Conference (CCWC)}.\hskip 1em plus 0.5em minus 0.4em\relax IEEE, 2018, pp.
  99--103.

\bibitem{johnson2016googles}
M.~Johnson, M.~Schuster, Q.~V. Le, M.~Krikun, Y.~Wu, Z.~Chen, N.~Thorat,
  F.~Viégas, M.~Wattenberg, G.~Corrado, M.~Hughes, and J.~Dean, ``Google's
  multilingual neural machine translation system: Enabling zero-shot
  translation,'' 2016.

\bibitem{snapchat}
\BIBentryALTinterwordspacing
Z.~Li, X.~Wang, X.~Lv, and T.~Yang, ``Sep-nets: Small and effective pattern
  networks,'' \emph{CoRR}, vol. abs/1706.03912, 2017. [Online]. Available:
  \url{http://arxiv.org/abs/1706.03912}
\BIBentrySTDinterwordspacing

\bibitem{He2015a}
K.~He, X.~Zhang, S.~Ren, and J.~Sun, ``Delving deep into rectifiers: Surpassing
  human-level performance on imagenet classification,'' in \emph{Proceedings of
  the IEEE international conference on computer vision}, 2015, pp. 1026--1034.

\bibitem{Zoph2018}
B.~Zoph, V.~Vasudevan, J.~Shlens, and Q.~V. Le, ``Learning transferable
  architectures for scalable image recognition,'' in \emph{Proceedings of the
  IEEE conference on computer vision and pattern recognition}, 2018, pp.
  8697--8710.

\bibitem{tensorflow:model-zoo}
\BIBentryALTinterwordspacing
``{Hosted models: TensorFlow Lite},'' Google Brain Team, [accessed Janurary
  2020]. [Online]. Available:
  \url{https://www.tensorflow.org/lite/guide/hosted\_models}
\BIBentrySTDinterwordspacing

\bibitem{Netravali2015}
R.~Netravali, A.~Sivaraman, S.~Das, A.~Goyal, K.~Winstein, J.~Mickens, and
  H.~Balakrishnan, ``Mahimahi: Accurate record-and-replay for {HTTP},'' in
  \emph{2015 {USENIX} Annual Technical Conference ({USENIX} {ATC} 15)}.\hskip
  1em plus 0.5em minus 0.4em\relax Santa Clara, CA: {USENIX} Association, Jul.
  2015, pp. 417--429.

\bibitem{He2015}
K.~He, X.~Zhang, S.~Ren, and J.~Sun, ``Deep residual learning for image
  recognition,'' in \emph{Proceedings of the IEEE conference on computer vision
  and pattern recognition (CVPR 2016)}, 2016.

\bibitem{Szegedy2016}
C.~Szegedy, V.~Vanhoucke, S.~Ioffe, J.~Shlens, and Z.~Wojna, ``Rethinking the
  inception architecture for computer vision,'' in \emph{Proceedings of the
  IEEE conference on computer vision and pattern recognition}, 2016, pp.
  2818--2826.

\bibitem{Bianco2018}
S.~Bianco, R.~Cadene, L.~Celona, and P.~Napoletano, ``Benchmark analysis of
  representative deep neural network architectures.''

\bibitem{guo2018cloud}
T.~Guo, ``Cloud-based or on-device: An empirical study of mobile deep
  inference,'' in \emph{2018 IEEE International Conference on Cloud Engineering
  (IC2E)}.\hskip 1em plus 0.5em minus 0.4em\relax IEEE, 2018, pp. 184--190.

\bibitem{Howard2017}
A.~G. Howard, M.~Zhu, B.~Chen, D.~Kalenichenko, W.~Wang, T.~Weyand,
  M.~Andreetto, and H.~Adam, ``Mobilenets: Efficient convolutional neural
  networks for mobile vision applications.''

\bibitem{tensorflow:lite}
\BIBentryALTinterwordspacing
``Tensorflow lite,'' Google Inc., [accessed October 2019]. [Online]. Available:
  \url{https://www.tensorflow.org/lite}
\BIBentrySTDinterwordspacing

\bibitem{caffe2}
\BIBentryALTinterwordspacing
{Facebook Research}, ``Caffe2,'' [accessed February 2019]. [Online]. Available:
  \url{https://caffe2.ai}
\BIBentrySTDinterwordspacing

\bibitem{pixel2_specs}
``{Pixel 2 - Wikipedia},'' \url{https://en.wikipedia.org/wiki/Pixel\_2}, 2019.

\bibitem{Ogden2018}
S.~S. Ogden and T.~Guo, ``{MODI}: Mobile deep inference made efficient by edge
  computing,'' in \emph{{USENIX} Workshop on Hot Topics in Edge Computing
  (HotEdge 18)}.\hskip 1em plus 0.5em minus 0.4em\relax Boston, MA: {USENIX}
  Association, Jul. 2018.

\bibitem{Crankshaw2017}
D.~Crankshaw, X.~Wang, G.~Zhou, M.~J. Franklin, J.~E. Gonzalez, and I.~Stoica,
  ``Clipper: A low-latency online prediction serving system,'' in \emph{14th
  USENIX Symposium on Networked Systems Design and Implementation (NSDI 17)},
  2017, pp. 613--627.

\bibitem{Olston2017}
C.~Olston, N.~Fiedel, K.~Gorovoy, J.~Harmsen, L.~Lao, F.~Li, V.~Rajashekhar,
  S.~Ramesh, and J.~Soyke, ``Tensorflow-serving: Flexible, high-performance ml
  serving.''

\bibitem{Gao:2018:LLR:3190508.3190541}
P.~Gao, L.~Yu, Y.~Wu, and J.~Li, ``Low latency rnn inference with cellular
  batching,'' in \emph{Proceedings of the Thirteenth EuroSys Conference}, ser.
  EuroSys '18.\hskip 1em plus 0.5em minus 0.4em\relax New York, NY, USA: ACM,
  2018.

\bibitem{LeMay2020}
M.~LeMay, S.~Li, and T.~Guo, ``{Perseus: Characterizing Performance and Cost of
  Multi-Tenant Serving for CNN Models},'' in \emph{2020 IEEE International
  Conference on Cloud Engineering (IC2E)}.\hskip 1em plus 0.5em minus
  0.4em\relax IEEE, 2020.

\bibitem{Satyanarayanan:2009:CVC:1638591.1638731}
M.~Satyanarayanan, P.~Bahl, R.~Caceres, and N.~Davies, ``The case for vm-based
  cloudlets in mobile computing,'' \emph{IEEE Pervasive Computing}, vol.~8,
  no.~4, Oct. 2009.

\bibitem{Kang2017}
Y.~Kang, J.~Hauswald, C.~Gao, A.~Rovinski, T.~Mudge, J.~Mars, and L.~Tang,
  ``Neurosurgeon: Collaborative intelligence between the cloud and mobile
  edge,'' in \emph{ACM SIGARCH Computer Architecture News}, vol.~45,
  no.~1.\hskip 1em plus 0.5em minus 0.4em\relax ACM, 2017, pp. 615--629.

\bibitem{Teerapittayanon2016}
S.~Teerapittayanon, B.~McDanel, and H.-T. Kung, ``Branchynet: Fast inference
  via early exiting from deep neural networks,'' in \emph{2016 23rd
  International Conference on Pattern Recognition (ICPR)}.\hskip 1em plus 0.5em
  minus 0.4em\relax IEEE, 2016, pp. 2464--2469.

\bibitem{Teerapittayanon2017}
S.~Teerapittayanon, B.~McDanel, and H.~T. Kung, ``Distributed deep neural
  networks over the cloud, the edge and end devices,'' in \emph{2017 IEEE 37th
  International Conference on Distributed Computing Systems (ICDCS)}.\hskip 1em
  plus 0.5em minus 0.4em\relax IEEE, 2017, pp. 328--339.

\bibitem{caffe:model-zoo}
``{Caffe: Model Zoo},'' \url{http://caffe.berkeleyvision.org/model\_zoo.html}.

\bibitem{Zhang2019}
C.~Zhang, M.~Yu, W.~Wang, and F.~Yan, ``{MArk: Exploiting Cloud Services for
  Cost-Effective, SLO-Aware Machine Learning Inference Serving},'' in
  \emph{2019 USENIX Annual Technical Conference (USENIX ATC'19)}, 2019.

\bibitem{Kannan2019}
R.~S. Kannan, L.~Subramanian, A.~Raju, J.~Ahn, J.~Mars, and L.~Tang,
  ``{GrandSLAm: Guaranteeing SLAs for jobs in microservices execution
  frameworks},'' in \emph{Proceedings of the Fourteenth EuroSys Conference
  2019}.\hskip 1em plus 0.5em minus 0.4em\relax ACM, 2019, p.~34.

\bibitem{Ousterhout2013}
K.~Ousterhout, P.~Wendell, M.~Zaharia, and I.~Stoica, ``Sparrow: distributed,
  low latency scheduling,'' in \emph{Proceedings of the Twenty-Fourth ACM
  Symposium on Operating Systems Principles}.\hskip 1em plus 0.5em minus
  0.4em\relax ACM, 2013, pp. 69--84.

\bibitem{Mitzenmacher2001}
M.~Mitzenmacher, ``The power of two choices in randomized load balancing,''
  \emph{IEEE Transactions on Parallel and Distributed Systems}, vol.~12,
  no.~10, pp. 1094--1104, 2001.

\bibitem{Bodik2010}
P.~Bodik, A.~Fox, M.~J. Franklin, M.~I. Jordan, and D.~A. Patterson,
  ``Characterizing, modeling, and generating workload spikes for stateful
  services,'' in \emph{Proceedings of the 1st ACM symposium on Cloud
  computing}.\hskip 1em plus 0.5em minus 0.4em\relax ACM, 2010, pp. 241--252.

\bibitem{Widmer1996}
G.~Widmer and M.~Kubat, ``Learning in the presence of concept drift and hidden
  contexts,'' \emph{Machine learning}, vol.~23, no.~1, pp. 69--101, 1996.

\bibitem{imagenet_cvpr09}
J.~Deng, W.~Dong, R.~Socher, L.-J. Li, K.~Li, and L.~Fei-Fei, ``{ImageNet: A
  Large-Scale Hierarchical Image Database},'' in \emph{CVPR09}, 2009.

\bibitem{Iandola2016}
F.~N. Iandola, S.~Han, M.~W. Moskewicz, K.~Ashraf, W.~J. Dally, and K.~Keutzer,
  ``{SqueezeNet: AlexNet-level accuracy with 50x fewer parameters and $<$0.5MB
  model size}.''

\bibitem{Simonyan2014}
K.~Simonyan and A.~Zisserman, ``Very deep convolutional networks for
  large-scale image recognition.''

\bibitem{ILSVRC15}
O.~Russakovsky, J.~Deng, H.~Su, J.~Krause, S.~Satheesh, S.~Ma, Z.~Huang,
  A.~Karpathy, A.~Khosla, M.~Bernstein, A.~C. Berg, and L.~Fei-Fei, ``{ImageNet
  Large Scale Visual Recognition Challenge},'' \emph{International Journal of
  Computer Vision (IJCV)}, vol. 115, no.~3, pp. 211--252, 2015.

\bibitem{Han2015}
S.~Han, H.~Mao, and W.~J. Dally, ``Deep compression: Compressing deep neural
  networks with pruning, trained quantization and huffman coding,'' \emph{arXiv
  preprint arXiv:1510.00149}, 2015.

\end{thebibliography}

\end{document}